\documentclass[preprint2]{aastex62}

\begin{document}

\newcommand{\kms}{\ensuremath{\mathrm{km}\,\mathrm{s}^{-1}}}
\newcommand{\galunits}{\ensuremath{\mathrm{km}\,\mathrm{s}^{-1}\,\mathrm{kpc}^{-1}}}
\newcommand{\galacc}{\ensuremath{\mathrm{km}^2\,\mathrm{s}^{-2}\,\mathrm{kpc}^{-1}}}
\newcommand{\MLsun}{\ensuremath{\mathrm{M}_{\sun}/\mathrm{L}_{\sun}}}
\newcommand{\Lsun}{\ensuremath{\mathrm{L}_{\sun}}}
\newcommand{\Msun}{\ensuremath{\mathrm{M}_{\sun}}}
\newcommand{\Ha}{\ensuremath{\mathrm{H}\alpha}}
\newcommand{\SFR}{\ensuremath{\mathit{SFR}}}
\newcommand{\aveSFR}{\ensuremath{\langle \mathit{SFR} \rangle}}
\newcommand{\sfrate}{\ensuremath{\mathrm{M}_{\sun}\,\mathrm{yr}^{-1}}}
\newcommand{\Aunits}{\ensuremath{\mathrm{M}_{\sun}\,\mathrm{km}^{-4}\,\mathrm{s}^{4}}}
\newcommand{\surfdens}{\ensuremath{\mathrm{M}_{\sun}\,\mathrm{pc}^{-2}}}
\newcommand{\voldens}{\ensuremath{\mathrm{M}_{\sun}\,\mathrm{pc}^{-3}}}
\newcommand{\gevcc}{\ensuremath{\mathrm{GeV}\,\mathrm{cm}^{-3}}}
\newcommand{\etal}{et al.}
\newcommand{\LCDM}{$\Lambda$CDM}
\newcommand{\ML}{\ensuremath{\Upsilon_*}}
\newcommand{\Mst}{\ensuremath{M_*}}
\newcommand{\Mg}{\ensuremath{M_g}}
\newcommand{\Mb}{\ensuremath{M_b}}
\newcommand{\gobs}{\ensuremath{\mathrm{g}_{\mathrm{obs}}}}
\newcommand{\gtot}{\ensuremath{\mathrm{g}_{\mathrm{tot}}}}
\newcommand{\gbar}{\ensuremath{\mathrm{g}_{\mathrm{bar}}}}
\newcommand{\azero}{\ensuremath{\mathrm{g}_{\dagger}}}

\title{The Imprint of Spiral Arms on the Galactic Rotation Curve}

\author{Stacy S. McGaugh}
\affil{Department of Astronomy, Case Western Reserve University, Cleveland, OH 44106}

\begin{abstract}
{We discuss a model for the Milky Way obtained by fitting the observed terminal velocities with the radial acceleration relation. 
The resulting stellar surface density profile departs from a smooth exponential disk, having bumps and wiggles that correspond to massive spiral arms.
These features are used to estimate the term for the logarithmic density gradient in the Jeans equation, which turn out to have exactly the right 
location and amplitude to reconcile the apparent discrepancy between the stellar rotation curve and that of the interstellar gas.
This model also predicts a gradually declining rotation curve outside the solar circle with slope $-1.7\;\galunits$, as subsequently observed.}
\end{abstract}


\section{Introduction}
\label{sec:intro}

{The study of Galactic structure has a long history. 
Considerable recent progress has been fueled by} large surveys
like Gaia \citep{Gaia}, RAVE \citep{RAVE},  APOGEE \citep{APOGEE}, and many others \citep[see][]{BHGreview}.
{Further information can be obtained by applying to the Milky Way knowledge gained from studies of external galaxies. }

Ideally, we would like to have a detailed map of the Galaxy in six dimensional phase space. 
Short of such an ideal, it is convenient to describe the bulk properties of the Galaxy in simpler terms.
The surface brightness profiles of spiral galaxies are often approximated as exponential disks.
This is a form of data compression, reducing all the complexities of the image of a galaxy containing
$\sim 10^{11}$ stars to just two parameters: a central surface brightness and a scale length \citep[e.g.,][]{F70}. 
The exponential disk provides a useful shorthand for the basic parameters of rotationally supported
disk galaxies, and for discussing the range of physical properties over which they exist \citep[e.g.,][]{BIM97,S2006,GAMA}. 
The extent to which this approximation is valid and sufficient depends on the application.

The exponential disk is often a sufficient approximation in photometric applications. This is less true for dynamics. 
There exists an analytic solution for the rotation curve of a {razor thin} exponential disk \citep{F70,BT} that is 
frequently utilized for convenience. However, deviations from a pure exponential that are modest in photometric terms
become more pronounced in dynamics: the exponential disk it is a rather poor approximation to the circular velocity curve of the 
gravitational potential of a real stellar disk \citep{Sellwood99}. The bumps and wiggles matter \citep{Sancisi04}.
 {The purpose of this paper is to move beyond the exponential disk approximation
to a numerically defined radial mass profile that includes bumps and wiggles.}

Rotationally supported disk galaxies obey tight scaling relations like the baryonic Tully-Fisher relation \citep{btforig,verhTF},
the central density relation \citep{CDR}, and the Radial Acceleration Relation \citep[RAR:][]{RAR,OneLaw,LiRAR}. 
The RAR is a correlation between the radial acceleration predicted by the observed distribution of stars and gas
and that which is observed. The RAR provides an example of the importance of properly treating the bumps and wiggles
through numerical solution of the Poisson equation, as the scatter in this relation increases when the exponential disk
approximation is misapplied \citep{wrongRAR}.

The Milky Way is a typical spiral galaxy, so presumably obeys the same relations as the other disk galaxies in the local universe.  
Here we take a first step to move beyond the oversimplification of a smooth exponential stellar disk by applying these scaling
relations to the Milky Way. Specifically, we apply the RAR to precise measurements of the terminal velocities to infer 
the stellar mass profile of the Galaxy. 
{This is as close as we can come to treating our own Galaxy in the same manner as we do external spiral galaxies.}

{An important advantage of this procedure is that it requires only empirically established knowledge.
In particular, the theoretical basis of the RAR is irrelevant. Whether it is caused by MOND \citep{milgrom83,Milgrom2016metoo}, 
or is natural in \LCDM\ \citep{KW2017,Ludlow2017,Navarro2017}, or not \citep{CJP2015,Milgromcomment2016,FKP2018,YongHAR}, or 
maybe sort of \citep{DCL2016,Desmond2017}, is neither here nor there for the purposes of this paper. 
All that matters here is that there exists a relation between the observed and baryon-predicted accelerations
in external spiral galaxies that has been calibrated empirically \citep{RAR,OneLaw,YongMDAR} without reference to or
dependence on any particular theory hypothesized to explain the missing mass problem. 
The critical if obvious assumption is that the Milky Way adheres to the same relation that has been established for other galaxies.}

In section \ref{sec:galdat} we discuss the Galactic data that we employ.
{In section \ref{sec:method} we describe the method by which the pattern of bumps and wiggles in 
the stellar mass distribution is inferred from the terminal velocities with the help of the RAR.} 
In section \ref{sec:nummod} we provide a numerical model for the Milky Way with updated Galactic constants.
In section \ref{sec:tests} we test this model with the slope it predicts beyond the solar circle, 
and investigate its impact on the structure of the stellar rotation curve.
Brief conclusions are offered in section \ref{sec:conc}.

\section{Galaxy Data}
\label{sec:galdat}

There is a large and rapidly growing list of constraints on Galactic structure \citep{BHGreview}. 
Here we take a parsimonious approach, comparing the rotation curve inferred from the terminal velocities observed
in the interstellar gas to that derived from stellar data. 
The overall agreement between stars and gas is good by conventional astronomical standards, 
{differing by 5 -- 15 \kms}. Statistically,
this modest difference is highly significant, {as the formal uncertainties are only 1 -- 2 \kms}. 

Geometric considerations make terminal velocities precise tracers of Galactic rotation interior to the solar circle \citep{BM}. 
The Milky Way is dynamically cold, with gas and young stars exhibiting velocity dispersions of order $\sim 10\;\kms$ \citep{ZS19}. 
This is a tiny fraction of the local rotation speed ($\sim 230\;\kms$): such orbits are very nearly circular.
{We make the traditional assumption that the terminal velocities are a good proxy for circular motion.
Doing so allows us to build a first order model for deviations of the stellar mass profile from a smooth exponential disk. 
There may of course be flows along the spiral arms; constructing a hydrodynamic model of those is beyond the scope of this work,
which is intended as a first approximation for the mass in spiral arms. This a necessary step toward such models,
which will presumably further perturb the pattern of density perturbations derived here. Since the local velocity field is cold, 
I expect these additional perturbations to be small.}

We utilize the terminal velocities observed in both the 21 cm line of atomic hydrogen and the CO lines of molecular gas.
Data for the atomic gas come from \citet{MGDQ4} for the fourth quadrant and \citet{MGDQ1} for the first quadrant. 
These are in excellent agreement with CO observations of molecular gas \citep{clemens,luna} over the same range of Galactic longitude.

In order to map the terminal velocities into a rotation curve, it is necessary to specify an absolute scale: 
our distance from the Galactic center, $R_0$, and the circular speed of the Local Standard of Rest (LSR), $\Theta_0$ \citep[e.g., ][]{McMillan2010}. 
Here we adopt the recent result of the \citet{GRAVITY} obtained from observing the periapsis passage
of a star around the Galaxy's central supermassive black hole. 
This provides the remarkably accurate measurement $R_0 = 8.122 \pm 0.031\;\mathrm{kpc}$.
In order to specify $\Theta_0$, we note that the proper motion of the radio source associated
with the central black hole, Sgr A$^*$ \citep[$6.379\;\mathrm{mas}\,\mathrm{yr}^{-1}$:][]{SgrAstar}, 
specifies a transverse speed of $245.58 \pm 1.32\;\kms$ for the sun for the adopted value of $R_0$.
The peculiar motion of the sun in the azimuthal direction with respect to the LSR accounts for a small part of the transverse speed;
adopting $V_{\sun} = 12.24 \pm 0.47\;\kms$ \citep{solarmotion} leads to a circular speed for the LSR of $233.34 \pm 1.40\; \kms$.

Stellar data provide an independent probe of the Galactic potential.
Recently, \citet{Eilers2019} combined information from Gaia \citep{Gaia}  
and APOGEE \citep{APOGEE} to estimate the Galactic rotation curve
from 5 to 25 kpc. They adopt the same value of $R_0$ but obtain a slightly lower $\Theta_0 = 229.0 \pm 0.2\;\kms$.
We will examine this point more closely in section \ref{sec:solarmotion}; for now note that a simple shift in the solar motion does not
suffice to reconcile the difference between the stellar and gas rotation curves.
The allowed amplitude of such a shift is too small, and does nothing to reconcile the difference in shape
{between the nearly flat, featureless stellar rotation curve of \citet{Eilers2019} and the rich structure
observed in the terminal velocities \citep{clemens,luna,MGDQ4,MGDQ1}}. 

\section{Method: Application of the RAR}
\label{sec:method}

{A strong empirical connection between the distribution of luminous and total mass in spiral galaxies has long been known \citep{vAS1986}.}   
Observed features in the azimuthally averaged luminosity profile --- bumps and wiggles
caused by features like spiral arms --- leave a corresponding imprint on the rotation curve \citep{Sancisi04}.
{That this connection is related to a ubiquitous acceleration scale in galaxies has also long been known
\citep{S1990,S1996,SV1998,M1999ASPC,MDacc}. A simple function maps between the rotation curves of external galaxies 
and their azimuthally averaged surface density profiles. This is the empirical Radial Acceleration Relation \citep[RAR:][]{RAR}.}


\begin{figure*}
\plotone{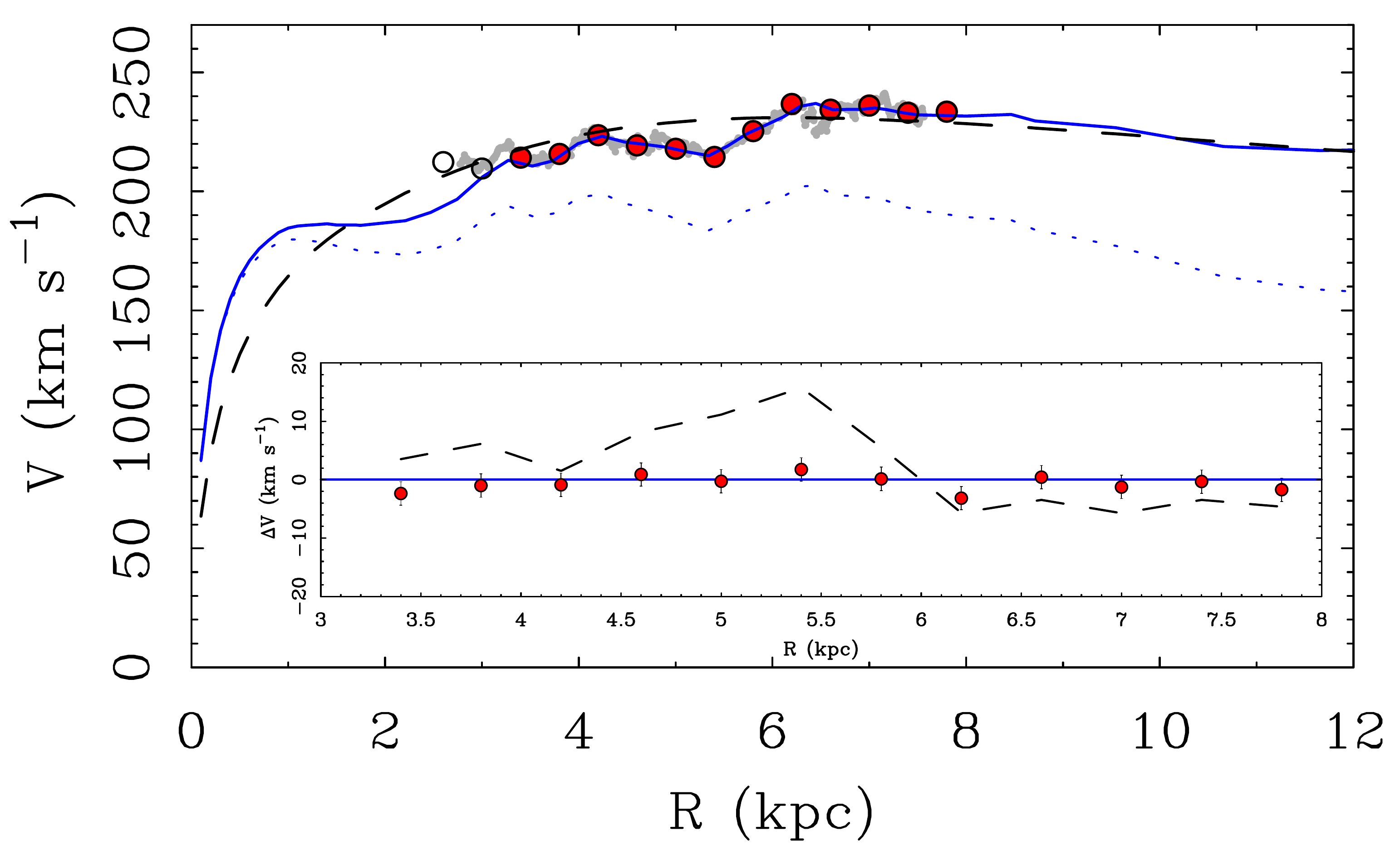}
\caption{The rotation curve of the Milky Way as indicated by terminal velocities in the fourth quadrant
in both CO \citep[][red points]{luna} and 21 cm \citep[][gray points]{MGDQ4}. 
The solid line that fits the $R > 3$ kpc data is the Q4MB model of \citet{M16}.
For comparison, the dashed line shows the  model of \citet{BovyRix}.
Residuals of the fit are shown in the inset.
The dotted line is the baryon-only rotation curve corresponding to the fitted rotation curve by way of the RAR.
This provides the run of stellar surface density in Table \ref{MWmassmodel} and Fig.\ \ref{fig:deriv}.
}
\label{fig:compare}
\end{figure*}

Quantitatively, the RAR is
\begin{equation}
\gobs = \frac{\gbar}{1-e^{-\sqrt{\gbar/\azero}}}
\label{eq:RFRfit}
\end{equation}
where $\gobs = V_c^2/R$ is the centripetal acceleration observed in a rotation curve,
$\gbar = (V_{bul}^2+V_{disk}^2+V_g^2)/R$ is the radial force generated by the gravitational potential of the baryons,
and $\azero = 3700\;\galacc$ is obtained from a fit to the data \citep{RAR}. 
The gravitational potential of the baryons is obtained from the observed, azimuthally averaged surface density profile
of stars and gas by numerically solving the Poisson equation. {In practice, this is accomplished with the program ROTMOD in the
GIPSY package \citep{GIPSY91}.} 

{Remarkably, the data for rotationally supported galaxies is consistent with a single, universal relation between
baryonic and observed acceleration \citep{SMmond,LivRev,LiRAR}. In external galaxies, both the surface brightness profile
(which leads to \gbar) and the rotation curve (which specifies \gobs) are measured. In the Milky Way, the terminal velocities
specify \gobs, but the baryon distribution is obscured by our location within the Galaxy. Here, we apply
the RAR obtained from external galaxies to the \gobs\ measured by the terminal velocities to infer \gbar\ in the Milky Way,
and the corresponding stellar mass distribution $\Sigma_*(R)$.}

This method was first suggested by \citet{M08} and applied in detail by \citet{M16}. In a nutshell,
{an initial guess is made for the baryonic mass distribution. This must include all relevant components:
bulge-bar, stellar disk, and gas.} We fix the mass models for the bulge-bar and gas to those adopted by \citet{M08},
scaling appropriately for the small difference in $R_0$. Though no specific attempt has been made to fit motions
in the region dominated by the bar, we will see that the triaxial bulge-bar model constructed
by \citet{M08} is in excellent agreement with the constraints of \citet{Portail2017}. 
{All detailed adjustments are made to the surface density profile of the stellar disk.}

Starting from a smooth exponential disk, we adjust the amount of stellar mass in each ring to improve the fit. 
The mass distribution of the stellar disk is iteratively adjusted
until a run of \gbar\ is obtained that maps through the RAR to fit the run of \gobs\ indicated by the terminal velocities.
See Fig.\ 1 of \citep{M16} for an illustration of the steps in the iteration.
In this fashion, a good fit to the CO data of \citet{luna} is obtained (Fig.\ \ref{fig:compare} and Table \ref{redchisq}).
The resulting baryonic mass profile is unique, but there 
is some limited freedom to partition mass between the various components so long as the overall baryonic surface density profile 
is maintained. Adopting a different bulge-bar model or 
gas profile would imply a trivial compensatory change to the profile of the stellar disk. 

{Fig.~\ref{fig:compare} compares one of the models obtained in this fashion by \citet{M16} with  a model from the literature.
The specific model illustrated (Q4MB) was fit to the fourth quarter terminal velocities assuming a best-guess value for the bulge-bar mass.
The comparison model is from \citet{BovyRix}. In contrast to the model Q4MB based on the \textit{radial} force, the stellar mass
model of \citet{BovyRix} was fit to the radial run of \textit{vertical} forces. 
The rotation curve illustrated in Fig.\ \ref{fig:compare} is obtained by applying the RAR to
the mass distribution found by \citet[][adopting their assumptions for bulge and gas]{BovyRix}.}

{The model of \citet{BovyRix} is typical of the field in that it assumes a smooth exponential disk and treats the scale length
as a parameter to be determined. It does a reasonably good job of matching the rotation curve, especially considering that
it was not fit to these data. Both the amplitude and shape are about right, and this level of consistency is typical of Galactic models.
Indeed, it is about as well as one can hope to do with a mass model composed of a single exponential disk \citep[see also][]{BinneyPiffl}.
Formally, the fit to the terminal velocities is terrible, with large $\chi^2$ (Table \ref{redchisq}). 
This is typical of exponential models for the Galactic disk that make no attempt to match the structure in the terminal velocity data.}

\begin{deluxetable}{ccccc}
\tablewidth{0pt}
\tablecaption{{Reduced $\chi^2_{\nu}$ \label{redchisq}}}
\tablehead{
& \multicolumn{2}{c}{All data} & \multicolumn{2}{c}{Excluding $44 < \ell < 55^{\circ}$} \\
\colhead{Model}  & \colhead{$V_c(R)$} & \colhead{$K_Z(R)$} & \colhead{$V_c(R)$} & \colhead{$K_Z(R)$} 
 }
\startdata
 BR13 & 14.35 & 0.75 & 6.06 & \dots \\
 Q4MB & \phn 0.60 & 1.69 & \dots & 1.02
\enddata
\tablecomments{BR13 from \citet{BovyRix}; Q4MB from \citet{M16}.}
\end{deluxetable}

We can check the Q4MB model, which was constructed to fit the radial force, to see what it predicts for the vertical forces
\citep[see Fig.~14 of ][]{M16}. The $\chi^2$ obtained in this fashion is acceptable (Table \ref{redchisq}), if not as good as
that of \citet{BovyRix}. This is to be expected, since there is no fitting to the vertical force $K_Z$:
this is just what the model predicts. However, as discussed
by \citet{M16}, the terminal velocities and the stellar data of \citet{BovyRix} do not sample exactly the same part of the Galaxy.
The maximum excursion is exactly where each model fits the data informing the other least well, between Galactic longitudes
$44 < \ell < 55^{\circ}$. We recalculate $\chi^2$ excising this region and the results for both models improve (Table \ref{redchisq}).
Though better, the smooth exponential model is still a poor description of the terminal velocities. In contrast, the model fit
to the terminal velocities does a good job of predicting the vertical forces with no further adjustment.
{This provides some measure of encouragement that we have obtained a reasonable description of the Galactic
stellar mass distribution that is a genuine improvement over a simple exponential disk.}

\section{Galactic Disk Model}
\label{sec:nummod}

Here we consider a numerical model for the stellar mass distribution of the Milky Way
in which $\Sigma_*(R)$ varies as implied by the RAR and the observed terminal velocities. 
Our starting point is the Q4MB model of \citet{M16} described above. 
{We make two minor modifications to the 2016 Q4MB model. 
The first is to update the Galactic size from $R_0 = 8$ to 8.122 kpc \citep{GRAVITY}.
This results in a larger mass for the Galaxy,
\begin{eqnarray}
M_* = 6.16  \times 10^{10}\;\mathrm{M}_{\sun} \\
M_{gas} = 1.22 \times 10^{10}\;\mathrm{M}_{\sun},
\end{eqnarray}
{up from $M_* = 5.57 \times 10^{10}\;\mathrm{M}_{\sun}$ for Q4MB.
This modest change takes the model from being very close to the stellar mass estimate of
\citet[$5.43 \pm 0.57 \times 10^{10}\;\mathrm{M}_{\sun}$]{McMillan2017} to being closer to that of 
\citet[$6.08 \pm 1.14 \times 10^{10}\;\mathrm{M}_{\sun}$]{LN2015}.
The stellar mass estimate here depends systematically on the calibration of the RAR,
the absolute scale of the Galaxy ($R_0, \Theta_0$), and its asymmetry. 
First and fourth quadrant models differ by $\sim 10\%$ in stellar mass \citep{M16},
which brings us to the second modification.
}

The second modification is a choice. Rather than fit each quadrant separately as in \citet{M16}, 
here we split the difference between the first and fourth quadrant terminal velocity data.
This provides an estimate of the equivalent azimuthally averaged mass distribution.
The difference between quadrants is modest, of order 10 \kms,
and has decreased with the recent first quadrant data of \citet{MGDQ1}. 
This small difference is nevertheless real, and appears to result 
from the non-axisymmetry in the Milky Way stemming from the pitch angles of spiral arms as they transition from one
quadrant to the next. Indeed, the bumps and wiggles inferred kinematically in each quadrant separately 
correspond well with known spiral arms \citep{M16}, {consistent with known constraints on
the morphology of the Milky Way \citep{dVP78,BHGreview}.}
However, it is beyond the scope of this paper to construct a model that accounts for azimuthal
as well as radial variations in the stellar surface density. {The azimuthally averaged variation is already a considerable step
beyond the exponential disk models typically considered in the literature (Fig.\ \ref{fig:compare}),}
and should be a reasonably good approximation in the limit of tightly wound spiral arms:
the typical pitch angles in the Milky Way are in the range $5 < \psi < 13^{\circ}$ \citep{HHS,Vallee17,BeSSeL}.
{Moreover, azimuthal averaging is the procedure that works in external galaxies \citep{Sancisi04}, and is how the RAR was constructed.}

The resulting mass model is given in Table~\ref{MWmassmodel}.
The first column is the Galactocentric radius for $R_0 = 8.122\;\mathrm{kpc}$.
The next three columns are the circular velocity of a test particle in the gravitational potential of each baryonic component:
bulge, gas, and stellar disk. The fifth column is the total circular velocity. The corresponding stellar\footnote{The bulge-bar  
model is triaxial \citep{M08} so the major axis cut of the surface density profile given in Table~\ref{MWmassmodel} does 
not correspond directly to the circular velocity of the equivalent axisymmetric mass distribution as it does for the disk components.}
and gaseous surface densities are provided in the last three columns.
{Note that $\Sigma_{disk}$ is the only quantity that has been adjusted to fit $V_c$.}

\begin{deluxetable}{rrrrrrrr}
\tablewidth{0pt}
\tablecaption{Milky Way Mass Model\label{MWmassmodel}}
\tablehead{
\colhead{R}  & \colhead{$V_{bul}$} & \colhead{$V_g$} & \colhead{$V_{disk}$} & \colhead{$V_c$}
 & \colhead{$\Sigma_{bul}$} & \colhead{$\Sigma_{disk}$} & \colhead{$\Sigma_g$} \\
 \colhead{kpc} & \multicolumn{4}{c}{$\kms$} & \multicolumn{3}{c}{$\surfdens$} 
 }
\startdata
  5.620 & 99.2 & 12.4 &174.7 &232.0 &    0.0 &198.0 &11.2 \\
  5.823 & 97.5 & 13.1 &180.3 &236.7 &    \dots &187.3 &10.6 \\
  6.027 & 95.8 & 11.6 &185.0 &240.5 &    \dots &171.2 &10.4 \\
  6.205 & 94.4 & 10.8 &189.0 &243.9 &    \dots &165.9 &10.6 \\
  6.384 & 93.1 & 12.1 &193.6 &247.9 &    \dots &145.2 &10.9 \\
  6.546 & 91.9 & 14.7 &194.3 &248.9 &    \dots & 95.5 &10.9 \\
  6.725 & 90.7 & 17.4 &190.4 &246.3 &    \dots & 86.4 &10.7 \\
  6.887 & 89.6 & 19.3 &188.5 &245.2 &    \dots & 82.2 &10.3 \\
  7.026 & 88.7 & 19.8 &187.6 &244.8 &    \dots & 80.1 & 9.6 \\
  7.164 & 87.9 & 17.9 &187.2 &244.5 &    \dots & 74.5 & 8.7 \\
  7.285 & 87.1 & 11.6 &186.6 &243.9 &    \dots & 65.1 & 8.2 \\
\enddata
\tablecomments{Table \ref{MWmassmodel} is published in its entirety in the electronic edition of the Astrophysical Journal.
A portion is shown here for guidance regarding its form and content.}
\end{deluxetable}

The model is fit to a rather small range of radii interior to the solar circle ($3 < R < R_0$), {and only that portion clearly beyond the
end of the bar ($R > 6$ kpc) is robust. This small range nevertheless provides enough of a handle for the RAR to place the Milky Way
in the context of other spiral galaxies, and to make predictions outside of the fitted range. For example,} 
the total velocity can be estimated beyond the solar circle by applying the RAR to the measured baryonic distribution.
Extrapolation to $R = 20$ kpc, roughly the edge of the HI disk, indicates an enclosed dynamical mass of $2.1 \times 10^{11}\;\Msun$. 
This compares well with independent estimatesat this radius: 
2.1 \citep{Kupper2015}, 1.9 \citep{PostiHelmi}, 2.2 \citep{Watkins2019}, and $2.14 \times 10^{11}\;\Msun$ \citep{MalhanIbata}.
The shape of the extrapolated rotation curve can now be tested in greater detail.

\section{Tests of the Model}
\label{sec:tests}

{The model employed here was first reported in \citet{M2018}, and has not been altered here.
It makes two \textit{a priori} predictions that can now be tested with data that have appeared since the initial publication of the model.
First, extrapolation of the model makes a specific prediction for the outer slope of the rotation curve. Second,
variations in the stellar density profile impact the term in the Jeans equation that 
depends on the derivative of the density gradient. This impact is distinctive as the derivative $\partial \ln \Sigma_*/\partial\ln R$
can be large and change suddenly: what appears to be a subtle feature in the luminosity profile may nevertheless
have an outsized impact on the circular velocity derived from the Jeans equation.
The test here is whether the stellar rotation curve becomes consistent with the terminal velocities
for the same pattern of bumps and wiggles in the stellar mass profile. }

\subsection{The Outer Slope of the Rotation Curve}
\label{sec:outer}

\begin{figure*}
\plotone{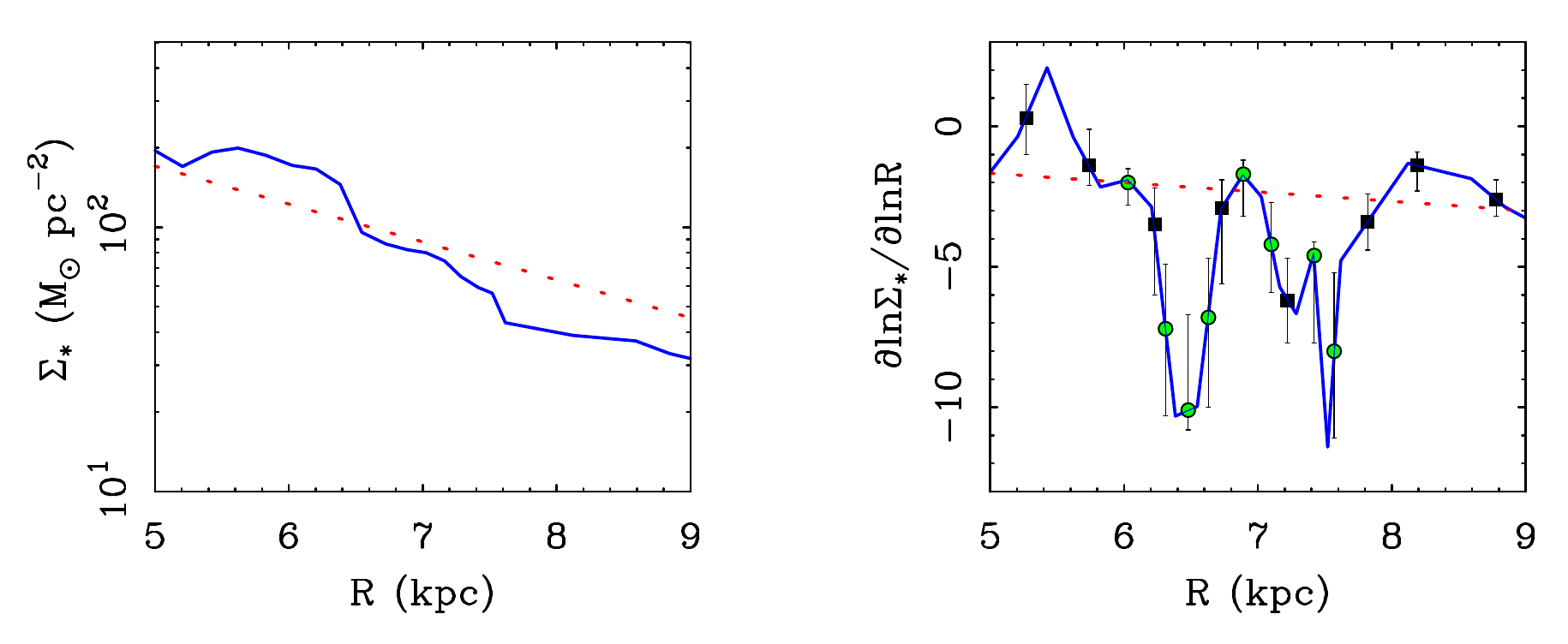}
\caption{The stellar surface density profile from Table \ref{MWmassmodel} (solid line, left) and the corresponding logarithmic
derivative (right, Table \ref{densityderiv}). An exponential disk with a 3 kpc scale length \citep[as employed by ][]{Eilers2019} is shown as the dotted line. 
The term for the logarithmic derivative obtained from the numerical model differs from that of the exponential approximation.
The model (solid line) is sampled for use in the Jeans equation in the same radial bins used by \citet{Eilers2019}, 
with the uncertainty estimated from the range of variation within each bin (points with error bars).
{We also analyze the data in smaller bins (green circles; Table \ref{densityderivhirez}) as a test of the effects of resolution on the method.}
}
\label{fig:deriv}
\end{figure*}

Interior to the solar circle, the bumps and wiggles corresponding to spiral arms are pronounced, and the slope of the rotation curve varies abruptly \citep{MQ2007,M16}.
Exterior to $R_0$, there is considerably less information from which to infer the presence of bumps and wiggles, though there are some hints \citep{SofueDip}.
The average rate of decline of the outer rotation curve can be predicted.
Over the range $9 < R < 19\;\mathrm{kpc}$, {the slope of the model is $dV/dR = -1.7 \pm 0.1\;\galunits$.}
Further out, the model continues to decline, albeit more gradually: $V(R = 100\;\mathrm{kpc}) \approx 190\;\kms$, or perhaps
a bit more ($\sim 196\;\kms$) if coronal gas \citep{coronabaryons} is included in the baryon budget.
This is a large extrapolation. How far such an extrapolation can be made depends on the range of validity of the RAR 
and the effects of external objects like the LMC \citep{Besla2010}.

One of the primary findings of \citet{Eilers2019} is that the rotation curve of the Milky Way is not perfectly flat, but declines slowly. 
This behavior is typical of bright galaxies \citep{URC,Noordetal}.
Quantitatively, \citet{Eilers2019} measure a rate of decline of $dV/dR = -1.7 \pm 0.1\;\galunits$. 
This is in reasonable agreement with the findings of \citet{Mroz}, who independently obtain $dV/dR = -1.34 \pm 0.21\;\galunits$ from Cepheids.
These results are in good agreement with the predictions of the model considered here, which preceded these data and has not been fit to them in any way.
The observed rate of decline in the Galactic rotation curve is a natural consequence of the RAR and the compact stellar mass distribution of the Milky Way.

The precise slope predicted by the model depends on the absolute scale of the Milky Way, 
with the exact value of $R_0$ being the largest source of systematic uncertainty.
To illustrate this effect, the first quadrant model of \citet[with $R_0 = 8$ kpc]{M16} 
has $dV/dR = -1.2\;\galunits$ and the fourth quadrant model has $dV/dR = -1.5\;\galunits$ over the corresponding radial range. 
These are obviously similar, but the relatively small change in $R_0$
from 8.0 to 8.122 kpc does have an effect on the extrapolated slope. 
Still greater changes to $R_0$ could lead to flat or even rising rotation curves, 
{and it would become difficult to find a tenable solution for $R_0 > 8.4$ kpc.}

{The declining slope of the model is not perfectly linear, though it is reasonably well approximated 
as such over the range $9 < R < 19\;\mathrm{kpc}$ (Fig.\ \ref{fig:MW2019_Gaia}).
Further out, the rotation speed continues to decline, albeit more gradually. 
From $R = 20$ to 40 kpc, the model predicts an average rate of decline of $-0.7\;\galunits$,
and from 40 to 100 kpc, the rate of decline slows to $-0.1\;\galunits$.
This is a large extrapolation. How far such an extrapolation can be made depends on the range of validity of the RAR 
and the effects of external objects like the LMC \citep{Besla2010} and M31.
Even the effects of coronal gas \citep{coronabaryons} may become relevant: should these be
included in the baryonic mass budget at $R \approx 100$ kpc?}

\subsection{The Impact of Variations in the Density Profile on the Circular Velocity Derived from the Jeans Equation}
\label{sec:jeans}

There is a modest but significant discrepancy between the rotation curve determined from stellar data by \citet{Eilers2019} 
and that indicated by the terminal velocity data over the range where the data overlap ($5 < R < R_0$).
In order to derive the circular velocity curve from stellar data, it is necessary to account asymmetric drift: 
{the population-dependent deviation of stellar motions from circular orbits \citep{BM}. 
This depends on the density profile of the tracer population. The model considered here has a unique pattern of bumps and wiggles
in its density profile that should leave a distinctive imprint on the predicted rotation curve. If this pattern is sufficiently close to correct,
it should reconcile the differences between the stellar rotation curve and the terminal velocities from which the model was inferred.
This would imply that the density variations are physical, with the corresponding gravitational effect.}

Accounting for asymmetric drift is accomplished with the Jeans equations \citep[][]{BT}.
Adopting the notation of \citet[their equation 4]{Eilers2019},
\begin{equation}
V_c^2(R) = \langle v_{\phi}^2 \rangle - \langle v_R^2 \rangle \left(1+\frac{\partial \ln \nu}{\partial \ln R}+\frac{\partial \ln \langle v_R^2 \rangle}{\partial \ln R} \right),
\label{eq:Jeans}
\end{equation}
where the circular velocity of a test particle ($V_c$) depends on both the azimuthal ($v_{\phi}$) and radial motions ($v_R$) averaged over many stars.   
It further depends on the logarithmic density gradient ($\partial \ln \nu/\partial \ln R$) and the logarithmic gradient 
in the square of the radial velocities ($\partial \ln \langle v_R^2 \rangle/\partial \ln R$). For the latter, we adopt the exponential 
fit\footnote{Close examination of the data in the left panel of Fig.\ 1 of \citet{Eilers2019} suggests a flattening in the radial 
velocity gradient slightly in excess of the fit beyond $R > 14$ kpc, so we tested a model with a break to a constant value beyond this radius. 
This makes no perceptible difference to the results.} of \citet[their Fig.\ 1]{Eilers2019}.  
The logarithmic density gradient is provided by our model, which replaces the smooth exponential disk assumed by \citet{Eilers2019}.

We numerically compute the logarithmic density gradient from the model in Table \ref{MWmassmodel}.
Neglecting vertical flaring, the variation in the three dimensional density $\nu$ is restricted to that in the surface density 
profile, so $\partial \ln \nu/\partial \ln R = \partial \ln \Sigma_*/\partial \ln R$. The stellar profile and its logarithmic derivative are 
shown in Fig.\ \ref{fig:deriv}. {We use this in place of a smooth exponential disk in equation \ref{eq:Jeans} to derive the circular velocity.} 

\begin{deluxetable}{rlc}
\tablewidth{0pt}
\tablecaption{\footnotesize{Jeans Term \& Rotation Curve \label{densityderiv}}}
\tablehead{
\colhead{Radius}  & \colhead{${\partial \ln \Sigma_*}/{\partial\ln R}$} & \colhead{$V_c$} \\
 \colhead{kpc} & & \colhead{\kms} 
 }
\startdata
 5.27 & \phs $0.3^{+1.2}_{-1.3}$ &$224.07^{+5.64}_{-5.26}$ \\
 5.74 &$-1.4^{+1.3}_{-0.7}$ &$234.54^{+2.94}_{-5.06}$ \\
 6.23 &$-3.5^{+1.3}_{-2.5}$ &$241.99^{+8.70}_{-4.80}$ \\
 6.73 &$-2.9^{+1.0}_{-2.7}$ &$237.84^{+9.14}_{-3.65}$ \\
 7.22 &$-6.2^{+1.5}_{-1.5}$ &$247.55^{+4.74}_{-4.81}$ \\
 7.82 &$-3.4^{+1.0}_{-1.0}$ &$238.08^{+3.14}_{-3.15}$ \\
 8.19 &$-1.4^{+0.5}_{-0.9}$ &$230.49^{+2.79}_{-1.70}$ \\
 8.78 &$-2.6^{+0.7}_{-0.6}$ &$231.25^{+1.95}_{-2.25}$ \\
 9.27 &$-3.2$               &$232.20^{+0.62}_{-0.72}$ \\
 9.76 &$-3.3$               &$231.76^{+0.52}_{-0.42}$ \\
10.26 &$-3.2$               &$230.82^{+0.40}_{-0.44}$ \\
10.75 &$-3.1$               &$229.26^{+0.41}_{-0.38}$ \\
11.25 &$-2.62$               &$227.10^{+0.54}_{-0.33}$ \\
11.75 &$-2.74$               &$226.95^{+0.39}_{-0.40}$ \\
12.25 &$-2.85$               &$225.32^{+0.37}_{-0.51}$ \\
12.74 &$-2.97$               &$223.86^{+0.46}_{-0.54}$ \\
13.23 &$-3.08$               &$224.04^{+0.40}_{-0.57}$ \\
13.74 &$-3.20$               &$220.57^{+0.51}_{-0.64}$ \\
14.24 &$-3.32$               &$220.44^{+0.66}_{-0.77}$ \\
14.74 &$-3.43$               &$220.77^{+0.68}_{-0.65}$ \\
15.22 &$-3.55$               &$220.27^{+0.80}_{-1.06}$ \\
15.74 &$-3.67$               &$220.62^{+1.07}_{-0.84}$ \\
16.24 &$-3.78$               &$219.41^{+1.48}_{-1.20}$ \\
16.74 &$-3.90$               &$215.79^{+1.43}_{-1.39}$ \\
17.25 &$-4.02$               &$219.77^{+1.85}_{-1.44}$ \\
17.75 &$-4.14$               &$217.07^{+1.65}_{-2.22}$ \\
18.24 &$-4.25$               &$211.24^{+1.88}_{-1.76}$ \\
18.74 &$-4.37$               &$213.02^{+2.77}_{-2.31}$ \\
19.22 &$-4.48$               &$209.88^{+2.36}_{-2.54}$ \\
19.71 &$-4.59$               &$205.34^{+2.26}_{-2.99}$ \\
20.27 &$-4.72$               &$203.30^{+2.89}_{-3.15}$ \\
20.78 &$-4.84$               &$201.66^{+3.37}_{-3.33}$ \\
21.24 &$-4.95$               &$198.81^{+6.50}_{-5.99}$ \\
21.80 &$-5.08$               &$217.43^{+12.18}_{-15.38}$ \\
22.14 &$-5.16$               &$180.36^{+18.57}_{-28.58}$ \\
22.73 &$-5.30$               &$196.76^{+19.05}_{-27.64}$ \\
23.66 &$-5.52$               &$180.19^{+16.74}_{-18.67}$ \\
24.82 &$-5.79$               &$202.33^{+6.12}_{-6.50}$ \\
\enddata
\end{deluxetable}

{The only difference between our analysis and that of \citet{Eilers2019} is the logarithmic density gradient.}
This is given in Table \ref{densityderiv} along with the resulting rotation curve.
This is a small but distinguishable adjustment to the rotation curve reported by \citet{Eilers2019}.

\begin{figure*}
\plotone{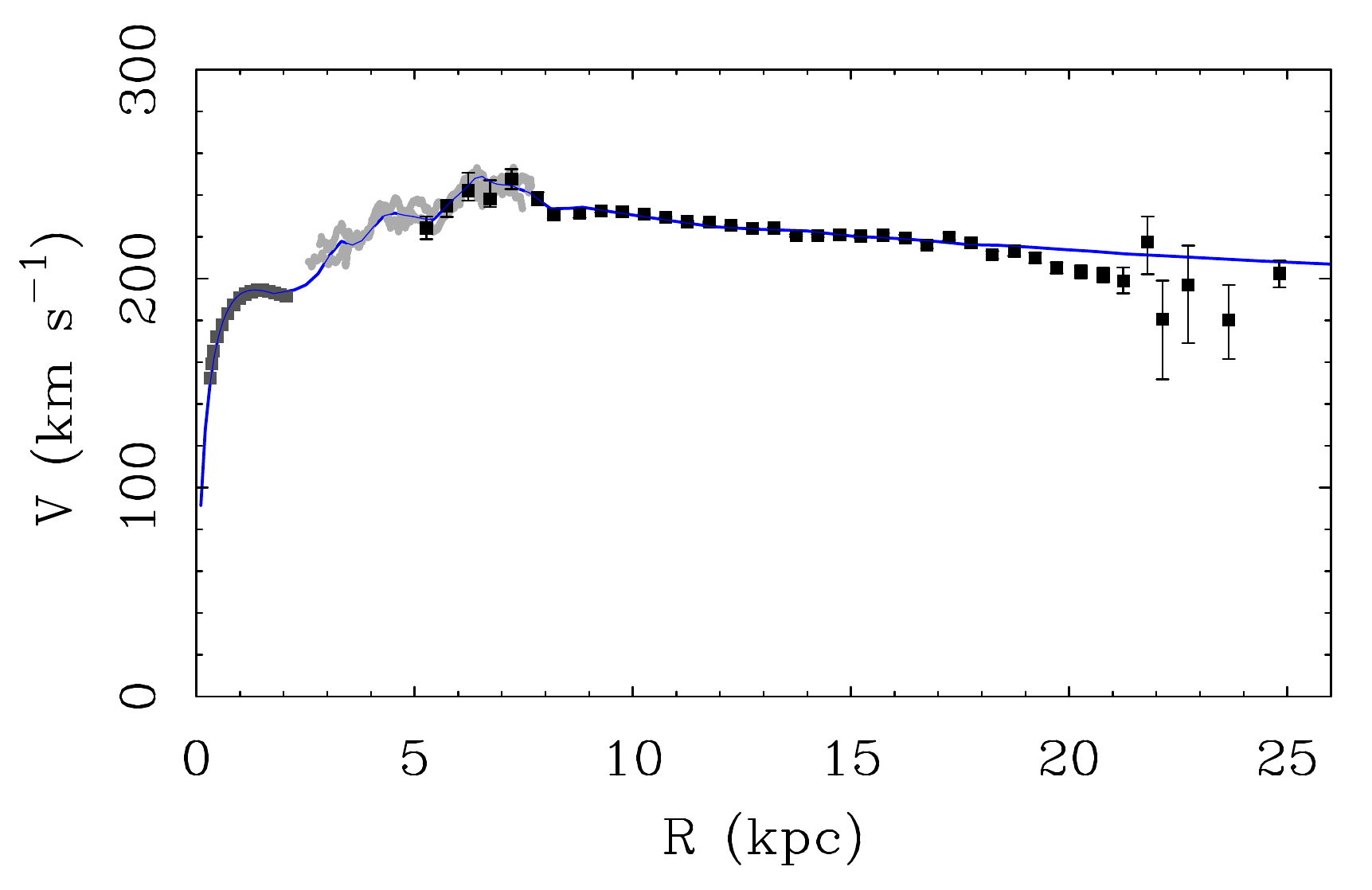}
\caption{The rotation curve of the Milky Way showing the model discussed here (blue line).
The light grey points are the terminal velocity data of \citet{MGDQ4,MGDQ1} to which the model was fit.
The model is also in good agreement with the circular velocity curve estimated 
for $R < 2.2\;\mathrm{kpc}$ by \citet[][dark gray squares]{Portail2017} to which it was not fit.
It is also in good agreement with the data of \citet[][squares with error bars]{Eilers2019} once the latter 
have been analyzed with the stellar density profile of the model (Fig.\ \ref{fig:deriv}).
}
\label{fig:MW2019_Gaia}
\end{figure*}

The logarithmic density gradient varies widely and sometimes abruptly in our numerical model. 
{This contrasts with the smooth variation of an exponential disk, and is the sole cause of the differences reported here.
Note that the absolute surface densities of the stellar disks in Fig.\ \ref{fig:compare} are not much different, but the derivatives often are.
These cause the bumps and wiggles to have an outsized impact on the application of the Jeans equation.}

{Radial bins} are chosen to match those of \citet{Eilers2019} {to facilitate direct comparison.}
We estimate the uncertainty in the logarithmic density gradient by the range of variation within each radial bin. 
{This provides a rough idea of the extent to which the central value of the bin is representative of the stars within it.}
These uncertainties are propagated into those in the circular velocity.
{There is considerable uncertainty in these uncertainties when the gradient varies abruptly.
Nevertheless, we do consider this an improvement over the usual practice or \textit{assuming} 
the smooth and gradual variation of a purely exponential disk. The derivative is an inherently local quantity that can
and apparently does vary substantially from point to point.}

{Indeed,} over the range where they can be estimated, the uncertainty in the logarithmic density gradient
is the dominant source of error, {yet seems not to have been considered previously. With the strictly}
monotonic density variation {of an exponential disk}, the formal errors are all $< 2\;\kms$.
After including this source of variation, the errors are in the range 2 --- 9 \kms, often exceeding the contribution of all potential
systematic errors discussed by \citet[see their Fig.\ 4]{Eilers2019}. This necessarily follows when we drop the artificial assumption
of a smooth exponential disk: variations in the scale length of the disk matter less than local variations in the derivative due to bumps and wiggles. 

It is only possible to estimate variations in the density profile interior to the solar circle where detailed information is provided by the terminal velocities.
Beyond the solar radius, we are obliged to assume a smoothly varying profile. Indeed, for $R > 11$ kpc,
the baryonic profile of the model discussed here is well approximated by an exponential with a scale length of 4.2 kpc, 
which we adopt for simplicity. We lack any information with which to estimate the uncertainty in the
logarithmic density gradient for $R > 9$ kpc, so these are not given in Table \ref{densityderiv}. 
The uncertainties in the circular velocity are simply carried over from \citet{Eilers2019} in this range.
{These may be understated, as we cannot preclude the presence of bumps and wiggles exterior to the solar radius.}

\begin{deluxetable}{rlc}
\tablewidth{0pt}
\tablecaption{\footnotesize{High Resolution Jeans Term \label{densityderivhirez}}}
\tablehead{
\colhead{Radius}  & \colhead{${\partial \ln \Sigma_*}/{\partial\ln R}$} & \colhead{$V_c$} \\
 \colhead{kpc} & & \colhead{\kms} 
 }
\startdata
6.03   &    $-2.0^{+0.5}_{-0.8}$  &$236.90^{+3.05}_{-2.06}$ \\
6.31	 &    $-7.2^{+2.3}_{-3.1}$  &$254.39^{+10.16}_{-7.56}$ \\
6.48   &   $-10.1^{+3.4}_{-0.7}$  &$262.87^{+2.39}_{-10.60}$ \\
6.63   &    $-6.8^{+2.1}_{-3.2}$  &$251.35^{+10.29}_{-6.80}$ \\
6.89   &    $-1.7^{+0.5}_{-1.5}$  &$233.27^{+5.14}_{-1.96}$ \\
7.10   &    $-4.2^{+1.5}_{-1.7}$  &$241.29^{+5.51}_{-4.89}$ \\
7.42   &    $-4.6^{+0.5}_{-3.1}$  &$242.30^{+9.60}_{-1.84}$ \\
7.57   &    $-8.0^{+2.8}_{-3.1}$  &$252.54^{+9.08}_{-8.22}$ \\
\enddata
\end{deluxetable}

Figure \ref{fig:MW2019_Gaia} shows our model, the terminal velocity data of \citet{MGDQ4,MGDQ1}, and the revised rotation curve 
in Table \ref{densityderiv}. Also shown is the circular velocity curve for $R < 2.2$ kpc from Fig.\ 23 of \citet{Portail2017} with
radii scaled to be consistent with the $R_0$ adopted here. The agreement between the model and the various data is excellent
{once variations in the density gradient are accounted for. Note that the model has only been fit to the terminal velocities from
$3 < R < R_0$. The match to the rest of the data is a successful prediction of a model that predated the data.}

\begin{figure*}
\plotone{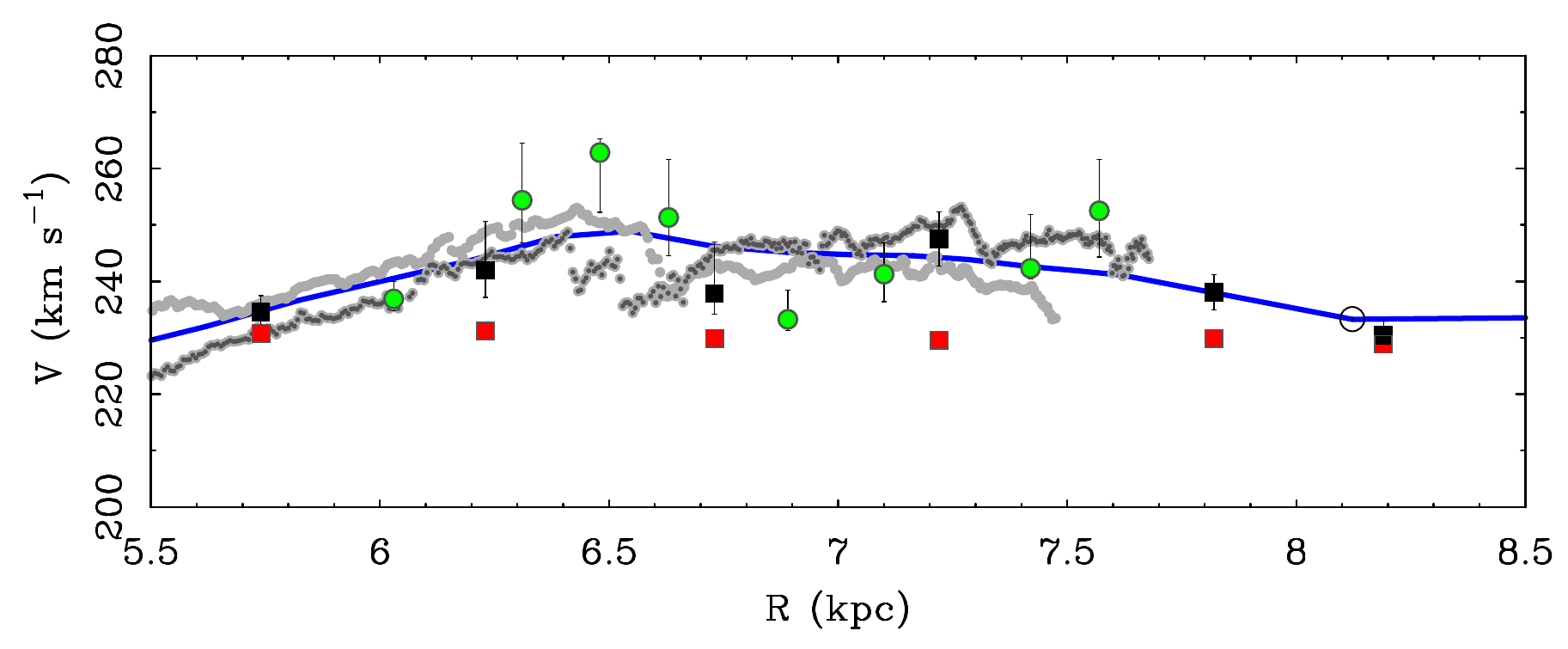}
\caption{A close examination of the rotation curve where variations in the surface density profile are most pronounced.
The open circle is the solar value (8.122 kpc, 233.3 \kms).
Light grey points are the first quadrant terminal velocity data \citep{MGDQ1}; dark grey points are the fourth quadrant
terminal velocity data \citep{MGDQ4}. The blue line that splits the difference between these two is the model described in the text
and seen in Fig.\ \ref{fig:MW2019_Gaia}. The red squares are the stellar rotation curve reported by \citet{Eilers2019}. 
The formal uncertainties on these data are less than the size as the squares, so the discrepancy from the terminal velocity data is highly significant. 
Black squares are these same data after consideration of variations in the logarithmic gradient term in the Jeans 
equation (Table \ref{densityderiv}). Green circles are more finely sampled bins of the same stellar data (Table \ref{densityderivhirez}).
Accounting for {point-to-point} variations in the density profile brings 
the stellar rotation curve into better agreement with the terminal velocity data.
}
\label{fig:MW2019_Gaia_zoom}
\end{figure*}

Of particular interest is that the bumps and wiggles in the stellar
surface density profile reconcile the rotation curve derived from the stellar data with that from the terminal velocities. 
The discrepancy between them {largely} vanishes. The model built to fit the terminal velocities correctly predicts the amount
of radial motion experienced by the stars.

{The shape of the density variations illustrated in Fig.\ \ref{fig:deriv} are imprinted on the rotation curve in Fig.\ \ref{fig:MW2019_Gaia_zoom}.
To explore this further}, we divide the data into finer bins in the region where the variation in 
the stellar surface density is greatest (Table \ref{densityderivhirez}). This allows us to sample some of the sharper features in $\Sigma_*(R)$,
at the expense of incurring larger uncertainties. This follows in part because there are fewer stars per bin, but more importantly because 
there are sometimes large variations in the gradient within the higher resolution bins. Nevertheless, the same result follows: the stellar
data become more consistent with the terminal velocities (Fig.\ \ref{fig:MW2019_Gaia_zoom}).
{Further progress likely requires self-consistent hydrodynamical modeling
in a non-axisymmetric disk.}

The {basic point here is that small scale} variations in the logarithmic density gradient {are non-negligible. 
Indeed, they appear to have about} the correct amplitude to reconcile stellar and interstellar data.
{A corollary is that a complete reconciliation \citep[][]{Bharat} is unlikely without consideration of this effect.}

\subsection{The Solar Motion}
\label{sec:solarmotion}

In constructing our model, we had adopted a circular speed of the LSR of $\Theta_0 = 233.3 \pm 1.4\;\kms$.
This is in good agreement with the value $\Theta_0 = 233.6 \pm 2.8\;\kms$ measured by \cite{Mroz}, 
but less so with the $\Theta_0 = 229.0 \pm 0.2\;\kms$ found by \citet{Eilers2019}. 
It is worth examining the role of the solar motion in this context.

The circular speed of the LSR found by \citet{Eilers2019} implies a solar motion of 
$V_{\sun} = 16.6\;\kms$ to be consistent with the proper motion of Sgr A$^*$ \citep{SgrAstar}.
This seems a bit large compared to more commonly obtained values \citep[$\sim 12\;\kms$:][]{BHGreview,ZS19}. 
We make our own estimate using the data of \citet{Eilers2019}, fitting the amplitude of the solar motion with the pattern
of bumps and wiggles fixed to that of our model.
Giving equal weight to all the data of \citet{Eilers2019}, we obtain 
\begin{equation}
V_{\sun} \approx 10.9\;\kms.
\label{eq:oursolarmotion}
\end{equation}
This is less than \citet{Eilers2019} find themselves, which occurs because of the adjustment in the structure of the rotation curve.
It is even a bit less than the 12.24 \kms\ \citep{solarmotion} that is built into our model.
This implies a slightly higher $\Theta_0 = 234.7\;\kms$, so in principle we could iterate to match both.
In practice, we are already there: the initial $\Theta_0 = 233.3$ is indistinguishable from $234.7\;\kms$ for any conceivable interpretation of the uncertainties, 
so there is no point in adjusting the model further. Even taken literally, this tiny difference more likely indicates 
a slight difference in the local pattern of bumps and wiggles than a change to the global Galactic constants. 
All that is needed is a slight change in the shape of the pattern near the sun \citep[e.g., ][]{SofueDip}, which is not well constrained by the data discussed here.
Such local features can certainly be relevant given the location of the sun just interior to the Orion spiral arm. 
{By the same token, such features can be missed by assuming they do not exist, as is implicitly done when adopting an exponential disk model.}

We make no attempt to place a formal uncertainty on the solar motion as the result is model specific. It also depends on the range of the data sampled.
The value in equation \ref{eq:oursolarmotion} gives the best overall agreement between the shape of model and that of the stellar rotation curve.
However, this varies by several \kms\ if we restrict the data to radii interior or exterior to the sun, or longitudes in the direction of motion or away from it.
This variation is not a random error, but a systematic uncertainty of $\sim 3\;\kms$. Indeed, the shape to which we are fitting is itself uncertain, 
depending on the pattern of bumps and wiggles within the solar radius and relying on an extrapolated density profile beyond it. 

The small absolute amplitude of the differences in the various estimates of the solar motion and circular speed of the LSR
likely render meaningless any further discussion, so we are content to have obtained values that are consistent with most previous efforts \citep{BHGreview,ZS19}.
For completeness, it seems worth noting that a closed orbit in a non-axisymmetric potential is an oval, not a circle \citep{BM}.
It is therefore conceivable that the tension seen between subsets of the data may be related to 
the subtle non-axisymmetry apparent in the terminal velocity data for different quadrants \citep{MGDQ4,MGDQ1}.
The Milky Way is a barred spiral galaxy, so we should not expect the closed orbit that the LSR notionally represents to be perfectly circular.

\section{Conclusions}
\label{sec:conc}

We have developed a model that describes the structure of the Galactic disk in detail beyond the simple approximation
of a smooth exponential disk. This model has a pattern of bumps and wiggles that correspond to massive spiral arms. 
These kinematically inferred spiral arms correspond well with the locations of photometrically known spiral arms \citep{M16}. 
That this should be the case is expected from experience with external galaxies \citep{Sancisi04}.

{Local variations in the radial density gradient due to spiral arms impact the rotation curve inferred with the Jeans equation.
The derivative is inherently a local quantity: how it varies on small scales is more important than the mean gradient of the scale length
averaged over the entire Galaxy. Accounting for local variations}
results in a model that fits the terminal velocity data in unprecedented detail, reconciles the apparent discrepancy
between the terminal velocities and independent stellar data, and accurately predicts the slope of the outer rotation curve. 

A salutary lesson is that we have reached a point where detailed numerical models for Galactic structure are necessary.
The commonly used exponential approximation is simply not adequate when confronted with modern data.
Moving beyond the exponential disk, as discussed here, is an obvious first step in modeling the in-plane dynamics of the Galaxy.
Further steps would be to build fully non-axisymmetric and non-stationary {multidimensional} models \citep[e.g.,][]{Monari2016,Monari2018}
{that account for azimuthal variations in stellar density and gas flows within spiral arms.}

\acknowledgements
I thank the referee for a thorough reading and stimulating exchange with a healthy perspective. 
Partial support for this work was provided by NASA through grant number HST-GO-15507.005 from the Space Telescope Science Institute, 
which is operated by AURA, Inc., under NASA contract NAS 5-26555.


\end{document}